\documentclass{llncs}
\usepackage{latexsym}
\usepackage{listings}
\usepackage{algorithmicx,algpseudocode}
\usepackage{tikz}
\usepackage{multirow}
\usepackage{color}
\usepackage{graphicx}
\usepackage{array}
\usepackage{url}
\usepackage{multirow}
\usepackage{amssymb}
\usepackage{comment}
\usepackage{rotating}
\usepackage{footnote}
\makesavenoteenv{tabular}

\lstdefinelanguage{APSL}{
  morekeywords={message, module, with, end, is, as, of, enum, record, union, tags, tagged, do, interactions,
                actor, import, on, otherwise, continue, next, where, from, init, state, or, anytime,
                type, codec, quit},
  sensitive=true,
  morecomment=[l]{\#},
  morestring=[b]",
}
\newcommand{\ttfy}[1]{{\small$\sf #1$}}
\newcommand{\NOTE}[1]{[[{\bf Note:} #1]]}
\renewcommand{\NOTE}[1]{}

\newenvironment{myquote}{\setlength{\leftmargini}{3mm}\quotation}{\endquotation}

\begin{document}

\mainmatter 

\title{Modeling and Testing Implementations of Protocols with Complex Messages}

\author{Tom Tervoort \and I.S.W.B. Prasetya}

\institute{Dept. of Inf. and Comp. Sciences,
  Utrecht University,
  the Netherlands \\
  \email{s.w.b.prasetya@uu.nl}
}

\maketitle

\vspace{-5mm}
\begin{abstract}
This paper presents a new language called APSL for formally describing protocols
to facilitate automated testing. 
Many real world communication protocols exchange messages whose structures are
not trivial, e.g. they may consist of multiple and nested fields, some could be optional,
and some may have values that depend on other fields. 
To properly test implementations of such a protocol, it is not sufficient to only
explore different orders of sending and receiving messages. 
We also need to investigate if the implementation indeed produces
correctly formatted messages, and if it responds correctly when it receives 
different variations of every message type. 
APSL's main contribution is its sublanguage that is expressive enough to
describe complex message formats, both text-based and binary.
As an example, this paper also presents a case study where APSL is used
to model and test a subset of Courier IMAP email server.
\end{abstract}

\vspace{-8mm}
\keywords{model based testing, testing protocols with complex messages, 
automated testing of protocols}

\section{Introduction} \label{sec.intro}

Modern communication protocols are often quite complex. Implementing one
is always tricky and error prone, and therefore an implementation should
be thoroughly tested before it is used. 
The complication lies not only in the interactions between the 
communicating parties, but also in the format of the messages. A message can be a quite
complex record structure with multiple fields, some could be optional, and some could
have delicate dependencies, which in turn are prone to errors.
%
%
Testing a protocol implementation can be made much more systematic if people first
construct a formal model of the protocol. In practice, people often do not do
this. They derive test cases directly from the natural language document that describes the protocol.
Although there are standards for such documents, e.g. RFC \cite{postel1997instructions}, and 
guides on how to write a good description \cite{scott1998guide}, a natural 
language description can still be deceiving and ambiguous. 

Attempts to provide formal languages to model protocols have been mostly focused on
describing the interaction parts, e.g. the languages SDL \cite{belina1989ccitt}, 
Estelle \cite{budkowski1987introduction}, and even UML \cite{lind2001specifying} (in particular the MSC part of UML).
A protocol is typically described as a system of participating actors, each is 
described by some form of labelled transition systems. 
However, these languages ignore the complexity of message formats, which make them not really 
usable for describing many real world protocols with complex messages. 
Model-based automated testing tools for protocols, e.g. 
TorX \cite{tretmans2003torx,goga2001comparing}, 
TGV \cite{fernandez1996using}, and Phact \cite{heerink2000formal}, follow the same trend.  
These tools can generate test cases from a formal model of 
a protocol, but they too focus on testing the interaction aspect of the protocol.
So, conformance validated by these tools does not imply that we have
covered all possible message structures. For example, we may still be left uncertain
whether an implementation under test would respond correctly if it gets a message 
with a certain optional field voided.

To address this gap, this paper contributes a language called 
APSL (A Protocol Specification Language)
to formally describe how the messages in a protocol are formatted and 
how parties in the protocol interact by exchanging these messages. 
APSL tries to be a light weight language while still be expressive enough to
describe real world text-based as well as binary protocols. 
A text-based protocol is a protocol whose messages are human readable strings,
whereas the messages of a binary protocol consist of low level bitstrings.
FTP, SMTP, and IMAP are examples text-based protocols. 
WebSocket is an example of a binary protocol.
%

From an APSL description of a protocol, an implementation of different parties in a protocol
can be automatically tested. Under the hood, APSL provides two basic testing functionalities:
(1) its generator can randomly generate messages of a {\em correct 
format}, and (2) its parser will check if the messages sent by an Implementation Under Test
(IUT) have a correct format. 
Just having these functionalities saves testers much effort
---else, they would have to write a custom message generator and parser
for each protocol they test, or to find a third party dedicated generator
and parser for the protocol. 
APSL's test engine is built on top those functionalities.
When invoked invoked, it will automatically traverse the IUT's interaction model
and in doing so test the IUT.
To provide flexibility, the engine can be parameterized by a traversal strategy to control how
the IUT is to be tested. 

This paper is structured as follows. Section \ref{sec.APSL} introduces APSL.
Due to limited space, we will only present a subset of APSL.
A description of its full syntax can be obtained  from \cite{APSLgit}.
Then, Section \ref{sec.testing}
explains how APSL test engine works. We have studied several examples to investigate the feasibility of using APSL for testing real world protocols. Section \ref{sec.experiment} will
highlight one of them: we show that we can use APSL to describe a subset of the IMAP protocol 
and test a server-side implementation of it. 
Section \ref{sec.relatedwork} discusses related work. Section \ref{sec.concl} concludes
and discusses some future work.

\section{Describing Protocols in APSL} \label{sec.APSL}

To describe a protocol in APSL we define a {\em message module} and an {\em interaction module}.
The first describes different types of messages that the protocol use, and
how they are formatted.  The second abstractly describes which {\em actors}  (parties)
take part in the protocol and how they interact by exchanging messages. 

\begin{figure}[t]
\begin{center}
\begin{tabular}{ccc}
\includegraphics[scale=0.7]{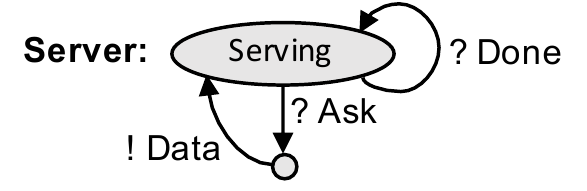} & \ \ \  &
\includegraphics[scale=0.7]{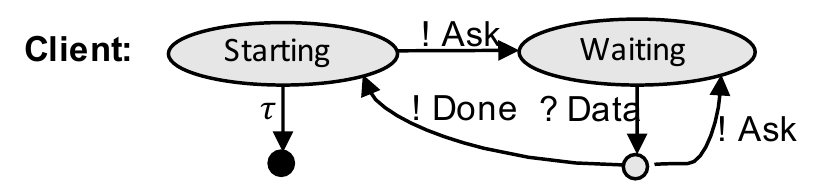}
\end{tabular}

\noindent
\begin{minipage}[t]{.6\textwidth}
\begin{lstlisting}[language=APSL,columns=fullflexible,frame=leftline]]
message module MyP
  message Ask   with h is Header(flag=1) end
  message Done  with h is Header(flag=3) end
  message Data  with ... 
  record Header with ...
end
\end{lstlisting}
\end{minipage}
\begin{minipage}[t]{.3\textwidth}
\begin{lstlisting}[language=APSL,columns=fullflexible,frame=leftline]]
interactions module MyP
  actor Client with ...
  actor Server with ...
end
\end{lstlisting}
\end{minipage}
\end{center}
\vspace{-5mm}
\caption{\ttfy{MyP} protocol, with two actors, a server and a client, and its top level APSL description.}
\label{fig.MyP.LTS}
\end{figure}

As a running example, consider a protocol called \ttfy{MyP} shown in Figure \ref{fig.MyP.LTS},
with a server and a client as actors, interacting as visually shown by
the corresponding Labelled Transition Systems (LTS's).
A transition in the LTS of an actor represents either an action by the actor
to send a message (denoted by '!'), 
or the receipt of a message (denoted by '?'), 
or a non-observable internal action by the actor (denoted by $\tau$).  
Each send and receive transition is labelled by
the type of the message that is to be sent or received. There are three types of messages
in the protocol.
'Data' messages are sent by the server to the client;
they carry payload. The client sends an 'Ask' message to the server to ask for an instance of 
Data, and a 'Done'  message if it decides that it has enough.

Figure \ref{fig.MyP.LTS} also shows the top level APSL description of \ttfy{MyP}.
The message module declares the above mentioned three types of messages,
whereas the interaction module declares the two actors (server and client).
Messages are basically records. A {\em record} is composed from {\em fields}, which comprise 
the core building blocks of an APSL specification. 
In the example, the message types \ttfy{Ask} and \ttfy{Done} are both defined to 
have a single field named \ttfy{h} of a type called \ttfy{Header}.

\subsection{Specifying complex messages}\label{subsec.AMSL}

Each field in a record has a name and a type, e.g. Integer
or Text. In typical specification languages, e.g. OCL \cite{cabot2012object} or Z \cite{smith2012object}, we do not have to care about how types are implemented. 
In protocol engineering we need to. Ultimately, messages are exchanged in bits. 
A protocol may insist on a specific way in which values are encoded in bits.
E.g. it may require a certain integer field to be represented in a 4-bits big-endian, 
while another integer field should be 32-bits, and so on. So
in addition to specifying the type of a field, in APSL we also need to specify a so-called {\em codec}
to describe how instances of the type should be formatted/represented in bitstrings.
The general syntax of a field declaration is:
\[ fieldname \ {\bf is} \ type \ {\bf as} \ codec \]
APSL comes with a range of common codecs. For example a codec called \ttfy{BoolBits} is used
to translate boolean $\sf true$ and $\sf false$ to specific bit patterns. 
The codec \ttfy{TerminatedText} can be used to format
a variable length text. It additionally appends a terminator to indicate where the
text ends.
Alternatively, if the length of the text is fixed, the codec \ttfy{FixedCountText}
can be used (when this codec is used, the receiver is thus assumed to know where in the received bitstring 
such a field ends). Almost all codecs have parameters to further specify
its format. E.g. \ttfy{BoolBits} requires us to specify the bit patterns
to use to represent \ttfy{true} and \ttfy{false}, whereas all text codecs such as \ttfy{FixedCountText}
require the used character set to be specified.

As an example, below we define the record type \ttfy{Header}, to be used 
in the \ttfy{Ask} and \ttfy{Done} messages of \ttfy{MyP}. We define it to be a record 
with two fields: 

\begin{lstlisting}[language=APSL,columns=fullflexible,xleftmargin=5mm,frame=leftline]
record Header with
  flag is Integer as BigEndian(signed=false,length=2)
  reserved is Binary(value=b'000000')
end
\end{lstlisting}

The field \ttfy{flag} is of type integer. The codec says that the integer will be represented 
by two bits in the unsigned big-endian format. The field \ttfy{reserved} is of type \ttfy{Binary}, 
which means it is simply a bitstring. For such, no codec is required. 

Types can be primitive (such as \ttfy{Integer}), records, or user defined types.
APSL also supports variant records (union types) ---see the documentation in \cite{APSLgit}.
APSL supports {\em dependent types}. These are types that
are parameterized by value-level expressions used to specify a subset of such a type.
For example, the type expression 
\ttfy{Binary(value{=}b'000000')} above specifies a subset consisting of a single value, 
namely the bitstring \ttfy{000000} 
(so, this is the only allowed value of the field \ttfy{reserved}).
As another example, \ttfy{Integer(min{=}0,max{=}500)} specifies a subset of 
integers, from 0 up to 500. 

An important principle in protocol design is that the receiver of a message should be able
to efficiently determine when a message ends, and similarly, when each field within a message
ends. One way to achieve this is to end a field with a specific bit pattern (e.g. as in
the \ttfy{TerminatedText} codec). Another common
convention used in various protocols is to have a field that specifies the length of the next field or fields. Dependent types are essential to capture such dependency. This is shown by the example below that defines a record type called \ttfy{DataItem}:

\begin{lstlisting}[language=APSL,columns=fullflexible,xleftmargin=5mm,frame=leftline]]
record DataItem with
  n  is Integer(min=0,max=500) as BigEndian(signed=false,length=32)
  data    is Binary(length=8*n)
  padding is Binary(length=8*(4 - n%4), char8_pattern=/\0*\1/)
end
\end{lstlisting}

A \ttfy{DataItem} basically carries some binary data. The length of this data is encoded in
the field \ttfy{n}, which should be an integer between 0 and 500. Its corresponding codec
specifies how this \ttfy{n} is encoded in bits. The field \ttfy{data} contains the actual 
data, its type parameter specifies that its length should be exactly \ttfy{8n} bits.
The field \ttfy{padding} is more complicated. Sometimes, a protocol requires that a message,
or a part of the message, to have a length which is a multiple of a certain number.
Suppose we want the total length of a \ttfy{DataItem} to be a multiple
of 32 bits. 
The field \ttfy{padding} is used to pad it if that was not the case. 
The type parameter \ttfy{length=8*(4 - n\%4)} specifies how long the padding should be.
The type \ttfy{Binary} can also be parameterized by a {\em regular expression} to specify allowed
bitstrings. Above, the regular expression in \ttfy{char8\_pattern=/\backslash0*\backslash1/} specifies 
that a bitstring of zero or more 0's closed by a 1 should be used as the padding.


For \ttfy{MyP}, we still have to define the message type \ttfy{Data}.  
The definition below shows an example of a more complicated message type involving
optional fields. Let's define the type \ttfy{Data} whose instances are records
that contain a header and an optional footer. The latter is specified by the type
\ttfy{Optional}, which is another example of a dependent type: the occurrence of the footer
is made to depend on the boolean value of another field, namely \ttfy{hasfoot}
(a footer exists of \ttfy{hasfoot} contains a value that represents $\sf true$, else
there is no footer). 

\begin{lstlisting}[language=APSL,columns=fullflexible,xleftmargin=5mm,frame=leftline,basicstyle=\sf\footnotesize]
message Data with
  h       is Header(flag=0)
  payload is List(elem=DataItem,max_length=4) 
          as CountPrefixList(count_codec=Word32Codec)
  hasfoot is Bool as BoolBits(falsehood_string=X'00',truth_string=X'ff')
  foot    is Optional(is_empty=!hasfooter,subject=Text) as ...  # some codec
end
\end{lstlisting}

Instances of \ttfy{Data} also contain a field called \ttfy{payload}, which is a
list of up to four instances of \ttfy{DataItem} (defined above) as the messages' payload.
APSL provides several standard codecs for translating a list (such as the field \ttfy{payload} above) 
to bitstrings. The codec \ttfy{CountPrefixList} used above will prepend
the list with an integer stating the length of the list and then sequentially write out
the items of the list to bitstring.

\subsection{Specifying interactions}

\begin{figure}[t]
\begin{lstlisting}[language=APSL,columns=fullflexible,xleftmargin=5mm,frame=leftline]
interactions module MyP
actor Client with
  init state Starting where anytime do send Ask next Waiting or do quit end
  state Waiting where on Data do send Ask continue 
                            or do send Done next Starting end
end
actor Server with
  init state Serving where on Ask do send Data continue
                         on Done  do continue end
end
\end{lstlisting}
\caption{The full interaction module of \ttfy{MyP}.}
\label{fig.MyP.AISL}
\end{figure}

In APSL, the interaction between the actors of a protocol is described in an interaction module,
by modeling each actor as an LTS a la \cite{tretmans1992formal} ---more precisely, as an input-output LTS (IOLTS) \cite{fernandez1996using}, which is an LTS where we distinguish between send and receive actions.
The IOLTS describes how the actor exchanges messages with its environment.
The latter represents, from the actor's perspective,  all the other actors.
The top level of the interaction module for
the example protocol \ttfy{MyP} was shown in Figure \ref{fig.MyP.LTS}.
Now, Figure \ref{fig.MyP.AISL} shows the full code
of this interaction module; it reflects the graphical IOLTS in Figure \ref{fig.MyP.LTS}.
A ${\tt send}\;M$ expression in an actor's description represents an action by the 
actor to send a message of type $M$ to the environment, whereas an
${\tt on}\;M$ expression represents the receipt of a message of type $M$ from the environment.
For example, the module in Figure \ref{fig.MyP.AISL}  states that the \ttfy{Client} has two named states, \ttfy{Starting} and \ttfy{Waiting}; the first
is also the initial state. When in the state \ttfy{Starting}, \ttfy{Client} can at any time
either quit, or send an \ttfy{Ask} message (an APSL model can thus be non-deterministic).
If it does the latter, it will also move to the
state \ttfy{Waiting}. If it receives a \ttfy{Data} message while in the state \ttfy{Waiting},
it will either send another \ttfy{Ask}, or send a \ttfy{Done} message, after which it moves
back to the \ttfy{Starting} state. 

Let's call the APSL description of an actor an {\em interaction model} of the actor.
Being a pure LTS, such a model only describes how the behavior of the actor depends on its LTS 
state and on the current message {\em type}. In particular, the model does not
specify what the content of the message should be, nor how the behavior of the actor
depends on this content. For example, when the \ttfy{Client} of \ttfy{MyP}
receives a \ttfy{Data} it can either respond with an \ttfy{Ask} or a \ttfy{Done}.
The decision could depend on the content of \ttfy{Data}. This is however abstracted away 
from the model in Figure \ref{fig.MyP.AISL}. To some extent, such dependency is still expressible
by an LTS, namely by defining different message types to represent
instances of \ttfy{Data} with different kinds of content. This works as long as the
number of the latter is finite.

\section{Testing} \label{sec.testing}

\begin{figure}[t]
\begin{center}
\includegraphics[scale=0.8]{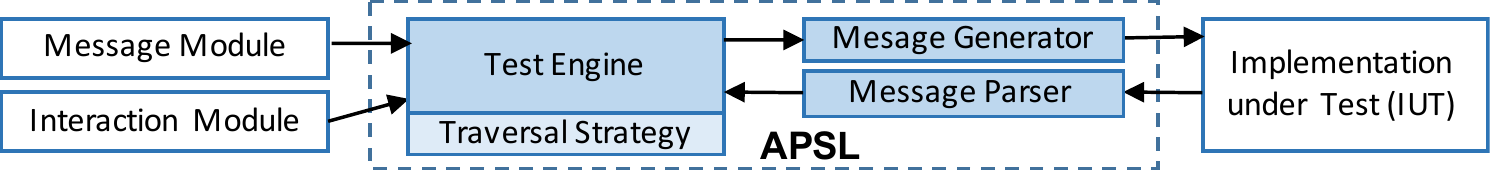}
\end{center}
\vspace{-5mm}
\caption{The architecture of APSL automated testing.}
\label{fig.architecture}
\end{figure}

Figure \ref{fig.architecture} shows the architecture of APSL automated testing framework.
Given a pair of message and interaction modules that describes a protocol, APSL test engine
can automatically test implementations of the protocol's actors. The actors are tested
one at a time. Consider an Implementation under Test (IUT) $I_A$ that implements an actor
$A$ in the interaction module. When invoked, the test engine generates a test in the form of 
a traversal over the IOLTS induced by $A$. So essentially, APSL does model-based
testing. The test engine takes the role
of $A$'s environment. A traversal starts at $A$'s initial state, and step-wisely extends itself
by following a transition to the next state. If the transition requires a message of some type $M$
to be sent to the IUT, the test engine will invoke the message generator to produce a 
random but correctly formatted message of type $M$, and send it to the IUT. Conversely, if the 
transition is triggered by the receipt of a message from the IUT, the message parser will
first check if the message has a valid format. 

In model-based testing it may seem sufficient to just generate model-level test cases.
However, note that to actually test an IUT we still need a message generator and a message parser 
to convert model-level test cases to concrete interactions with the IUT. These components are 
tedious and error prone to write. Fortunately, they are built in APSL, without which
testers will have to write them themselves for each protocol they test. 

\begin{figure}[t] 
\algblockdefx[NAME]{BlockBegin}{BlockEnd}%
[2][Unknown]{#1 #2}%
{{\bf end}}

\begin{algorithmic}[1]
\Procedure{\sf test}{$A:Actor$} \Comment{{\color{blue}the top level procedure to test an actor}}
    \State ${\sf traverse}(A,\; \tau^*(\{ \; \mbox{$A$'s initial state} \; \}),\;[\;])$ \Comment{{\color{blue}generate a traversal}}
\EndProcedure    

\State

\Procedure{\sf traverse}{$A,S,trace$}  \Comment{{\color{blue}$S$ is the set of possible current states}}
    \State \label{line.termination} {\bf if} some end condition {\bf then} {\bf throw Done}
    \State $status \gets {\sf parse}({\sf receive}())$
          
    \BlockBegin[{{\bf case}}]{$status$ {\bf of}}
      \State ${\sf ValidFormat} \ : \  msgType \gets {\sf getMsgTypeOf}({\sf getMsg}(status))$
      \State \label{line.invalidformat} ${\sf InvalidFormat} \ : \ {\bf throw\; invalidFormatError} \vspace{1mm}$
      \State ${\sf TimeOut}$ \  :  \ 
                 \Comment{{\color{blue}the IUT did not send anything}}
                  \BlockBegin[{{\bf do}}]{} 
                  \State $\rhd$ {\color{blue}calculate which types of messages can currently be sent to the IUT:}
                  \State $V \gets \{ \; \alpha \; | \;
                                              s{\in}S \wedge
                                              (\exists t.\; s \stackrel{?\alpha}{\rightarrow} t) \; \}$
                        
                  \State                          
                      ${\bf if}\; V{=}\emptyset \; {\bf then}\; {\sf traverse}(A,S)$
                  \State $\rhd$ {\color{blue}invoke a 'strategy' to decide which message type to send:}
                  \State \label{line.strategt} $msgType \gets strategy(S,V)$
                  \State $\rhd$ {\color{blue}generate an instance of the message type, and send it to IUT:}
                  \State ${\sf send}({\sf generateMsgInstance}(msgType))$
                  \BlockEnd
    \BlockEnd
    
    \State $trace \gets trace.{\sf addLast}(msgType)$   \Comment{{\color{blue}extend the trace}}
    
    \State $S \gets \{ \; t \; | \; s{\in}S \; \wedge \; s\stackrel{msgType}{\longrightarrow} t \; \}$ 
       \Comment{{\color{blue}calculate the possible next states}}
    \State \label{line.invalidtrace} ${\bf if} \; S{=}\emptyset \; {\bf then}\; {\bf throw\; invalidTraceError}(trace)$
    
    \State ${\sf traverse}(A,\tau^*(S),trace)$ \Comment{{\color{blue}recursively extend the traversal}}
\EndProcedure
\end{algorithmic}    
\caption{\em APSL's traversal algorithm. In the algorithm, $s \stackrel{a}{\rightarrow} t$
means that the actor $A$ contains a transition from the state $s$ to $t$, labelled with $a$.
The expression $\tau^*(S)$ means the set of
states that can be reached through zero or more $\tau$-transitions from some state in $S$.
This set always includes $S$ itself.} \label{fig.traversal.alg}
\end{figure} 

Figure \ref{fig.traversal.alg} shows APSL's traversal algorithm. The algorithm keeps track of
the trace of the types of the messages exchanged with the IUT so far. It also keeps track of
the actor's current state, so that it knows which transitions can be traversed to get to 
the next state. Since an APSL actor can be non-deterministic, the algorithm may
not be able to uniquely determine IUT's current state. It therefore maintains a set
$S$ of of possible current states. If $\sigma$ is a trace of message types,
let $s \stackrel{\sigma}{\Rightarrow} t$ means that there exists a path in the actor's LTS, to go from the state $s$
to the state $t$, such that if we remove all the $\tau$ labels from the trace of labels induced by this
path, the trace is equal to $\sigma$. Let $s_0$ be the actor's initial state, and suppose
$S$ is the set of current states according to the traversal algorithm, and $trace$ is
the traversal's trace so far. The algorithm maintains
the following property of $S$:

\[ S \ = \ \{ \; t \; | \; s_0 \stackrel{trace}{\Longrightarrow} t \; \} \]

If no error is found, line \ref{line.termination} decides when the traversal is 
ended, e.g. it could impose a limit of $N$ transitions.
When in the current states there are multiple receive transitions possible (note that these
are thus transitions where the test engine can send a message to the IUT), the algorithm
invoke a 'strategy' to chose which one to follow (line \ref{line.strategt}).
The default strategy simply randomly chooses a transition, thus inducing a random
traversal over the LTS. APSL provides a hook that allows this strategy to be redefined.

During a traversal, the test engine checks for two correctness aspects: (1) the IUT should produce
messages that conform with the format defined in the message module (line \ref{line.invalidformat}), 
and (2) the trace of the messages exchanged with the IUT should conform with  $A$ 
(line \ref{line.invalidtrace}). The latter means, more precisely, that:
\[ \{ \; t \; | \; s_0 \; \stackrel{trace}{\Longrightarrow} \; t \; \} \ \not= \ \emptyset \]
The engine itself will not send a message that would violate the trace conformance, so if it is violated,
the IUT is to blame.

At the end of a traversal, the test engine reports two kinds of coverage: 
the coverage over the transitions IUT's LTS model, and the coverage over the messages.
%
%
For the latter, the engine checks if every field (of every message type and 
the underlying record types) and every value of every enumeration type
has been tested.
Recall that fields are declared in a message and record definitions. In addition,
APSL also supports variant records and enumeration type (not shown in the paper).
A field is considered as covered by a traversal if it ever occurs in 
some message during the traversal. For an optional field, its absence also counts a coverage goal.

%


\section{Case Study} \label{sec.experiment}

To investigate whether APSL is expressive enough to describe and provide at least
basic automated testing of implementations of real world protocols we conducted two case studies:
Courier email server \cite{CourierIMAP} and 
Autobahn WebSocket server \cite{AutobahnWebsocket}.
On a smaller scale,  we have also used APSL to describe NTP time format \cite{NTP}, 
BSON data format \cite{bson}, and DNS message format \cite{DNSmsg}.
%
%

This section will present selected highlights of the email server case study. 
The full account of the case study, as well as that of the Autobahn WebSocket case study, can be found
in \cite{TervoortThesis}. Other mentioned examples can be found in APSL's home \cite{APSLgit}.
A Courier email server is used to store emails. 
Clients can connect to it to access and manage the stored emails.
The server implements the Internet Message Access Protocol (IMAP) protocol,
more precisely version 4rev1 defined in RFC 3501 \cite{crispinrfc}, that defines
how an IMAP server should communicate with its clients.
The following subset of the protocol is considered in the case study:

\begin{itemize}    
    \item The client can log-in with a username and password.
    
    \item In reply, the server greets the client. There are three different greetings, to indicate 
    that the user is either required to log in, or has already been authenticated, or  has no access.
    
    \item The client can examine, create, delete and rename mailboxes.
    
    \item The client can select a mailbox, after which emails within can be fetched, altered or copied. 
    Entire emails can be fetched, or only their metadata, or only their size and flags
    (e.g. whether an email has been read).
     The client can close the mailbox, allowing the client to continue opening another.
\end{itemize}
In this case study we want to know whether APSL is able to describe the part of the IMAP's server side protocol that covers the above functionalities. 
Section \ref{sec.model.IMAP} below shows that the answer is "yes".
Then, we then want to know if APSL can actually be used to test the Courier server;
this is discussed in Section \ref{subsec.casestudy.testing}.

\subsection{Modeling IMAP} \label{sec.model.IMAP}

\subsubsection*{Client commands}

Within IMAP, messages sent from the client to the server are called {\em commands} and messages from the server
are called {\em responses}.
A command starts with a tag, an arbitrary string of alphanumeric characters chosen by the client, followed by the 
command name, and the command's arguments whose types depend on the command. These parts are separated by a single space, and the command always ends with a line terminator. Since the first two fields are the same for every command, we can describe them as follows in APSL:

\begin{myquote}
\begin{lstlisting}[language=APSL,columns=fullflexible,xleftmargin=5mm,frame=leftline,basicstyle=\sf\footnotesize]
record CommandStart(commandName) with
  tag  is Tag                           as SpaceTerminated
  name is Identifier(value=commandName) as SpaceTerminated
end   
\end{lstlisting}
\end{myquote}
APSL supports user defined types and codecs. \ttfy{Tag} and \ttfy{Identifier} above 
are user defined types, whereas \ttfy{SpaceTerminated}
is a user defined codec. Their definitions:

\begin{myquote}
\begin{lstlisting}[language=APSL,columns=fullflexible,xleftmargin=5mm,frame=leftline,basicstyle=\sf\footnotesize]
type Identifier is Text(charset='ascii', pattern=/[!-~]+/, 
                     exclude_pattern=/ |\r\n|\*/,max_count=20)
type Tag is Identifier(pattern=/[0-9a-zA-Z]+/)
codec SpaceTerminated is TerminatedText(encoding='ascii',terminator=' ')
\end{lstlisting}
\end{myquote}

An \ttfy{Identifier} is thus a text
of maximum 20 characters which should not contain any space, line break, nor the '*' character, whereas a \ttfy{Tag}
is a purely alpha-numeric \ttfy{Identifier}.
The \ttfy{SpaceTerminated} codec formats a given text by appending a single space to its end.

The are 11 commands that correspond to the previously listed IMAP functionalities. They all start
with a \ttfy{CommandStart} followed by one or more command specific arguments. For example,
the structures of the command to delete a mailbox is shown below:
%
\begin{myquote}
\begin{lstlisting}[language=APSL,columns=fullflexible,xleftmargin=5mm,frame=leftline,basicstyle=\sf\footnotesize]
message DeleteCmd with
  _start  is CommandStart(command='DELETE')
  mailbox is MailboxId as LineTerminated
end
\end{lstlisting}
\end{myquote}
The last field in a command must be terminated by a line break. This is specified by the 
\ttfy{LineTerminated} codex, whose definition is analogous to \ttfy{SpaceTerminated}.


The field \ttfy{mailbox} in the \ttfy{DeleteCommand} poses however a challenge for automated testing.
It contains a value of the type \ttfy{MailboxId}, which is text representing the name of an
{\em existing} mailbox. APSL's message generator will have a very hard time generating such a name.
Any generic generator will have the same problem. Fortunately, this can be mitigated
in APSL over-specifying the type \ttfy{MailboxId}, e.g. as follows:
$\sf {\bf type} \;MailboxId \; {\bf is} \; Identifier(pattern{=}/INBOX|NOBOX/)$.
%


\subsubsection*{Server responses}

After receiving a command, the server answers with a tagged status response to indicate 
whether the commanded operation succeeds. This status response may however be preceded
by a number of untagged responses providing some relevant information depending on the 
command. A {\em tagged status response} starts with a tag, followed by a status text 
indicating the result of the triggering command, and another text with 
additional information. There are three types of statuses to indicate:
(1) the received command was successful, (2) the command failed,
and (3) the command was incorrectly formatted. 
As before, fields should be separated by a single space; the last one should end with a line break.
The formalization in APSL is shown below. 
We capture the three types of status responses with three different message types,
each is just a wrapper over the underlying \ttfy{StatusResponse} record type
that contains the actual fields mentioned above.

\begin{myquote}
\begin{lstlisting}[language=APSL,columns=fullflexible,xleftmargin=5mm,frame=leftline,basicstyle=\sf\footnotesize]
message OkResp  with resp is StatusResponse(response=ok) end
message NoResp  with resp is StatusResponse(response=no) end
message BadResp with resp is StatusResponse(response=bad) end
record StatusResponse(response) with
  tag  is Tag as SpaceTerminated
  _id  is StatusResponseId(value=response) as SpaceTerminated
  text is Text(charset='ascii', pattern=/[ -~]*/, exclude_pattern=/\r\n/) 
      as LineTerminated
end
enum StatusResponseId of Text with ok as 'OK'  no as 'NO'  bad as 'BAD' ... end
\end{lstlisting}
\end{myquote}

There are actually two additional statuses not shown above: \ttfy{PREAUTH}
(a server greeting indicating the client is already authenticated)
and \ttfy{BYE} (sent when logging out). These are used in certain untagged
responses. 

%
%

As said, the server may precede a tagged status response with a series of untagged responses.
There are two types of untagged responses: those that also report some status, and those 
that do not. Untagged status response has the same structure as a tagged response, except
that the tag is replaced by a '*'. It may report an intermediate status of an operation.
In addition to \ttfy{OK} or \ttfy{NO} statuses, some may report \ttfy{PREAUTH}
or \ttfy{BYE} statuses.
Non-status untagged responses always start with a message number (since they always relate 
to a specific e-mail), followed by an identifier indicating their type. 
They are expressed by wrapping a message type around the underlying record type below:

\begin{myquote}
\begin{lstlisting}[language=APSL,columns=fullflexible,xleftmargin=5mm,frame=leftline,basicstyle=\sf\footnotesize]
record UntaggedResponse(kind) with
  _asterisk is Text(value='*') as SpaceTerminated
  msg_id is MessageId as TextInteger(text_codec=SpaceTerminated)
  response_type is Identifier(value=kind) as SpaceTerminated
  info is Text(charset='ascii', pattern=/[ -~]*/, exclude_pattern=/\r\n/) 
       as LineTerminated
end
\end{lstlisting}
\end{myquote}

\vspace{-7mm}
\subsubsection*{The interaction}

The server-side protocol works as follows. Upon establishing a connection, the initial state is called
the \ttfy{ServerGreeting} state, where the server will then send a greeting to the client
in the form of an untagged response. The response can be: (1) an \ttfy{OK}, after which the server moves to the 
\ttfy{NotAuthenticated} state; (2) a \ttfy{PREAUTH} response, after which the server moves to the \ttfy{Authenticated} state; or a \ttfy{BYE} after which the server closes the connection. 

In the \ttfy{NotAuthenticated} state, the server can receive a request from the client to use a specific
authentication method. In this case study we only consider a simple \ttfy{LOGIN} command 
with a username and password as the authentication method; when it succeeds, the server moves to the \ttfy{Authenticated} state.
In the \ttfy{Authenticated} state, the server can receive commands that relate to the management of mailboxes. 
For example after receiving a \ttfy{SELECT} command, and if the command is successful,
the server moves to another state where commands such as storing and fetching emails are possible. 
Capturing those interactions in APSL is straightforward. As an illustration, below we show how transitions from the states \ttfy{ServerGreeting} and \ttfy{Authenticated} are modeled in APSL. The full interaction model, consisting of 7 states and 31 transitions, can be found in \cite{APSLgit}.

 
\begin{myquote}
\begin{lstlisting}[language=APSL,columns=fullflexible,xleftmargin=5mm,frame=leftline,basicstyle=\sf\footnotesize]
actor IMAPServer with    
  init state ServerGreeting where 
    anytime do send UntaggedOknext NotAuthenticated
          or do send PreAuthGreeting next Authenticated or do send Bye quit
  end 
  state Authenticated where
    anytime do send UntaggedOk continue
    on SelectCmd do next Selecting or do send NoResp continue
    on CreateCmd do send OkResp continue or do send NoResp continue
    on RenameCmd do send OkResp continue or do send NoResp continue
    ...
  end ...
\end{lstlisting}
\end{myquote}

\subsection{The testing} \label{subsec.casestudy.testing}

After the relevant part of IMAP is modeled in APSL, as discussed 
in Section \ref{sec.model.IMAP}, we can proceed to testing. 
We tested the Courier IMAP server \cite{CourierIMAP} version 4.10.0
as the IUT of the \ttfy{IMAPServer} model from Section \ref{sec.model.IMAP}.
The server was configured to have a mailbox already created, populated
with some emails. Since logging-in successfully to the server would problematical
for APSL's message generator, we over-specified the username and password
fields as we did to \ttfy{MailboxId}.
In general, a communication channel is needed to connect APSL test engine to an IUT.
In this case, a TCP/IP channel is needed, but this is already delivered by APSL 
itself. No further adapter was needed to facilitate testing.
The test engine was configured to keep traversing the IUT up to a certain
maximum number of steps $T_{max}$. 
As the test strategy, we simply used the default strategy that 
randomly chooses the next transition to follow if there are more than one possible.
This turned out to be sufficient for cover all feasible transitions.
Some transitions turned out to be unfeasible in the case study, e.g. the Courier server
never actually sends the \ttfy{Bye} transition (see the partial definition
of \ttfy{IMAPServer} actor in Section \ref{sec.model.IMAP}) even though this is allowed 
by the protocol.

Covering all (reachable) messages' fields and instances of enumeration types 
was more challenging, but still possible, with large enough $T_{max}$. 
For example, the command to fetch emails
has three variations: to fetch the emails' whole content, or only their envelope,
or only their flags. This is represented appropriately within the corresponding field
of the fetch-command message by specifying it using an enumeration type. We do not however make
the distinction visible at the transition level. That is, in our actor model
sending a fetch-command is represented by one transition rather than a choice of
three transitions. 
The latter is expressible in APSL, but would make the actor
model overly verbose and hence undesirable. Representing them as a single transition
however means that the traversal strategy in Figure \ref{fig.traversal.alg}
also needs to control the message generator so specify which variant of a 
message type it wants to generate. Such control is currently not possible; this is
future work.
 
No invalid format nor invalid trace error was found during the testing. Manual inspection on
the coverage reports of some of the produced traversals did reveal unexpected behavior.
When the test engine sends a command to the email server to delete a mailbox named \ttfy{INBOX}, 
the server refuses, which is indicated by sending a \ttfy{NoResp} response. 
This is correct, since IMAP's specification does not allow
\ttfy{INBOX} to be deleted. However, the investigation revealed that subsequent commands to select \ttfy{INBOX} is then rejected
(the server replies with a \ttfy{NoResp}). While this is allowed by the interaction model that we constructed, this is obviously a bug. 
With respect to the model, such a bug cannot be revealed by APSL, 
nor by any other purely LTS-based testing, since it is neither an invalid format nor an invalid trace error.
Refining the model where the special status of \ttfy{INBOX} is made explicit as a special message type
would reveal the error, but on the other hand would also make the model much more verbose, which is undesirable. A better way to deal with this would be to extend APSL so that transitions can 
be decorated by post-conditions. This is future work.


\section{Related Work} \label{sec.relatedwork}

The most used language for describing complex messages in communication protocols is probably Abstract 
Syntax Notation One (ASN.1) \cite{ASN.1}. It was introduced in 1984 and has since then underwent several
revisions. It is a joint standard of ISO, IEC, and ITU. ASN.1 is comparable with AMSL without codecs. 
Translations to the machine representation (the codec part) is expressed in a different language; there are some options, e.g. 
the Basic Encoding Rules (BER) notation, the XML Encoding Rules (XER) notation, or ITU's own Encoding Control Notation (ECN).
However ASN.1 has has grown to be rather large and complex: for example, it has more than 10 different string types, 
many intended for legacy character encodings. It also has ten expansive standards \cite{ATNexpansion}.
These make it more expensive to build an ASN.1-based tool.  Eclipse Titan \cite{Titan} is an example of an
ASN.1-enabled testing tool. The tool was originally developed by Ericsson in 2000, but has now been released as an open source
project. 
In Titan, test cases care abstractly
written in TTCN-3 \cite{grabowski2003introduction}. They will be compiled to concrete and executable tests. Messages in the test cases 
can refer to ASN.1 definitions. Despite its size (over 1.5 million lines of code), Titan is not
a model based testing (MBT) tool. In theory, it is possible to integrate e.g. an existing MBT tool 
with Titan, but we have yet to see such an integration. As far as we know, there is no MBT tool 
that is ASN.1 enabled out of the box.

In contrast to ASN.1, APSL is a simple language, but it is expressive enough to model complex messages. 
It already contains MBT tools out of the box. Typical testers only need to use the MBT tool. Its
test engine also allows a custom traversal strategy to be hooked, if the default random strategy is not sufficient. 
Tools builders may wish to  modify the test engine itself. The source code of APSL consists of less than 4000 lines of Haskell code, which should be feasible to work with. 

In addition to ASN.1, there are indeed more recent protocol description languages, e.g. 
Google Protocol Buffers \cite{GoogleProtocolBuffer} and Apache Thrift's interface description language
\cite{ApacheThrift}. Both allow for definitions of protocol messages in terms of simple data structures, 
and code can be generated that operates on the described data.
However, these systems use a particular binary encoding method. Therefore, the languages can not be used to 
describe protocols not originally designed with them. There is also the binpac language \cite{pang2006binpac}
which can be used to generate parsers for binary messages. A binpac definition of messages types are compiled 
to parsers in C{+}{+}. Pieces of additional C{+}{+} code can be mixed in to describe additional parser logic.
However, the tool set can only produce parsers. It does not allow for the construction of messages 
according to the definition (which is essential for a testing tool).

APSL relies on random generators to generate messages. Generating valid messages turns out to be, perhaps unsurprisingly, easy.
The most complex form of constraints on a field is a regular expression.
Such an expression can be constructively turned into generators. 
In comparison, the problem is analogous to generating valid input combinations in 
automated unit testing of functions or methods. 
However, functions' inputs are often constrained by a first order formula pre-condition
that can be hard for a random based testing tool such as QuickCheck \cite{claessen2011quickcheck} or T3 \cite{prasetya2013t3} to solve. If have to deal with a protocol whose messages are constrained in the similar way, we will face the same problem.
Fortunately, there are ways to at least mitigate the problem, e.g. by using a theorem prover like Z3 \cite{de2008z3} to solve the constraint, or, as in \cite{fraser2011evosuite,vos2013fittest}, by employing a search algorithm \cite{mcminn2004search}.

So far we assume the IUT to be a black box. In reality, we may have its source code which we can instrument. In particular,
we can insert code to collect runtime information.
Even if we do not have the source code, some bytecode can still be instrumented e.g. using tools like
ASM \cite{ASM} or Asil \cite{middelkoop2011functional}.
By collecting such information we can obtained more in-depth information about a test. E.g. if the IUT ever throws
an internal exception, and if so, which messages triggered it. We can also use the information to mine common
patterns which in turn can be used as anomaly detectors. E.g. it may be revealed that whenever the IUT can either 
send $a$ or $b$, it seems to always do the first whenever a certain internal function $f$ returns a 0.
An anomaly detector monitors a program at the runtime
and will raise a warning if the current behavior violates the behavior pattern it is keyed in. 
Note that this does not immediately mean that the behavior is erroneous; a human must investigate to decide this.
A tool like Daikon can be used to mine from runtime data \cite{ernst2007daikon}. More precisely,
it mines 'invariants' (properties over program variables that hold on certain points in the program).
Alternatively, techniques from data mining such as clustering can also be tried 
\cite{pol2015clustering}.



\section{Conclusion and Future Work} \label{sec.concl}

We conclude that APSL is able to express real life protocols with complex messages
such as IMAP and WebSocket and to provide at least basic automated 
model-based testing of these protocols. APSL has a test engine that can be parameterized
with a traversal strategy. The default random traversal strategy was enough
for the email server case study to cover all (feasible) transitions and message 
fields' variants. However, other protocols may be more challenging, thus 
requiring more powerful traversal strategies to be
written. There are approaches such as combinatorial testing \cite{nie2011survey}, search-based testing \cite{mcminn2004search},
or active learning \cite{bauersfeld2012reinforcement} which can be used as a base for such strategies. It requires however
further study to know what would work best on which classes of protocols.

The email server case study also shows a draw back of pure LTS model-based
testing. The server turned out to contain a real bug which a pure LTS-based approach would 
be blind without resorting to capturing certain value-level dependencies as
separate message types. This can be done, but it will lead to an overly verbose model
which is undesired. A better approach would be to extend the LTS-based approach
by allowing transitions to be decorated with pre- and post-conditions. This is future work.

APSL's current test engine can only test synchronous protocols. Some protocols, such as WebSocket,
are however asynchronous. An actor of such a protocol can buffer incoming messages and is thus not required
to respond immediately to an incoming message. Although we did a case study with a WebSocket server
\cite{TervoortThesis}, we only explored the synchronous behavior
of the server by forcing the test engine to wait long enough before it sends the next message to 
the IUT, and thus we can assume that by that time the server would have consumed all messages in its input buffer.
To also test the server's asynchronous behavior the test engine needs to be extended.
The challenge of asynchronous testing is that it introduces another aspect of non-determinism \cite{tretmans1992formal} that inhibits the test engine's ability to infer the state or states 
where the IUT might currently be; the number of possibilities can grow exponentially. 
Although this is a well known phenomenon in asynchronous testing, further study is needed to come up with strategies to keep this explosion
 manageable.


\vspace{-3mm}
\bibliographystyle{splncs03}
\bibliography{thebibs}

\end{document}